\documentclass[11pt]{article} 

\usepackage{deauthor,times,graphicx} 
\usepackage{tcolorbox}
\usepackage{amsfonts}
\usepackage{xpatch}
\usepackage{hyperref}
\usepackage{xspace}   %%%% 2025-01-09 newly added

\usepackage{caption}

\newcommand{\ours}{\textsc{Taiji}\xspace}

\makeatletter

\begin{document}

\title{\ours: MCP-based Multi-Modal Data Analytics on Data Lakes}
\author{Chao Zhang, Shaolei Zhang, Quehuan Liu, Sibei Chen, Tong Li, Ju Fan* \\ \small Renmin University of China\\  \small \{fanj@ruc.edu.cn\}}

\maketitle

\begin{abstract}
The {variety} of data in data lakes presents significant challenges for data analytics, as data scientists must simultaneously analyze multi-modal data, including structured, semi-structured, and unstructured data. While Large Language Models (LLMs) have demonstrated promising capabilities, they still remain inadequate for multi-modal data analytics in terms of accuracy, efficiency, and freshness.
First, current natural language (NL) or SQL-like query languages may struggle to precisely and comprehensively capture users’ analytical intent.
Second, relying on a single unified LLM to process diverse data modalities often leads to substantial inference overhead.
Third, data stored in data lakes may be incomplete or outdated, making it essential to integrate external open-domain knowledge to generate timely and relevant analytics results.
%
%The ``Variety'' of data poses significant challenges on data analytics in data lakes, which requires data scientists to simultaneously analyze the multi-modal data including structured, semi-structured, and unstructured data. Despite the emergent abilities of Large Language Models (LLMs), they still fall short of multi-modal data analytics regarding accuracy, efficiency, and freshness. First, existing Natural Language (NL) or SQL-like query languages can hardly deliver the users' query intent precisely and comprehensively. Second, leveraging one unified LLM to analyze the different kinds of data may incur significant inference overhead. Third, user data in the data lake may be incomplete and obsolete over time, and it is crucial to incorporate the open-domain knowledge for producing the up-to-date insights.
%
In this paper, we envision a new multi-modal data analytics system to address the aforementioned limitations. Specifically, we propose a novel architecture built upon the Model Context Protocol (MCP), an emerging paradigm that enables LLMs to collaborate with knowledgeable agents.
First, we define a semantic operator hierarchy tailored for querying multi-modal data in data lakes and develop an AI-agent–powered NL2Operator translator to bridge user intent and analytical execution.
Next, we introduce an MCP-based execution framework, in which each MCP server hosts specialized foundation models optimized for specific data modalities. This design enhances both accuracy and efficiency, while supporting high scalability through modular deployment.
Finally, we propose a updating mechanism by harnessing the deep research and machine unlearning  techniques to refresh the data lakes and LLM knowledges, with the goal of balancing the data freshness and inference efficiency.

%In this paper, we envision a new multi-modal data analytical system to overcome the above-mentioned limitations. Particularly, we design a new architecture based on Model Context Protocol (MCP), which is an emerging technique that can empower LLM with knowledgeable agents. Specifically, we first define a hierarchy of semantic operators upon multi-modal data lake and develop an NL2Operator translator using AI agent. We then propose an MCP-based operator execution architecture where each MCP server can host tailored foundation models for data of specific modality, thereby enhancing the accuracy and efficiency while delivering high scalability.  Next, we propose a updating mechanism by harnessing the deep research and machine unlearning  techniques to refresh the data lakes and LLM knowledges, with the goal of balancing the data freshness and inference efficiency.

%Next, we propose an adaptive model refreshing mechanism by harnessing modern database logging techniques to feed the latest data updates for the up-streaming MCP servers, with the goal of balancing the data freshness and inference efficiency.

% how to update the text/image LLM with record-based log entries?
% Deep Search for Data Augmentation

\end{abstract}

\section{Introduction}
\label{sec:introduction}

Modern data-intensive applications continuously generate diverse types of data stored in data lakes~\cite{nargesian2019data}, encompassing structured data (e.g., tables, graphs), semi-structured data (e.g., JSON, HTML), and unstructured data (e.g., text, images, video). This diversity introduces a critical challenge of \emph{data variety}~\cite{lu2019multi}, underscoring the need for integrated analysis across \emph{multi-modal datasets}, which often contain complementary information essential for extracting timely and valuable insights.
For example, in healthcare scenarios, physicians simultaneously analyze heterogeneous patient data, such as X-ray images and textual diagnostic reports, to support accurate and timely diagnoses. In e-commerce, users seek products by jointly exploring visual content, textual descriptions, and structured metadata.

% 1. Multi-Modal Data Analytics in Data Lake. Add An example
%Modern data-intensive applications have been producing various kinds of data residing in the data lake~\cite{nargesian2019data}, including structured data (e.g., tables, graphs), semi-structured data (e.g., JSON, HTML), and unstructured data (e.g., text, images, and video). Consequently, there is a pressing need to address the data ``Variety'' challenge~\cite{lu2019multi} and simultaneously analyze these multi-modal datasets because they contain enormous value and can deliver real-time insights. For instance, doctors in the hospitals need to analyze diverse patient datasets for timely disease diagnosis, including X-ray images and textual report; E-commerce users would like to search for ideal products from pictures, description and metadata.

Recently, the rapid advancement of Large Language Models (LLMs) has opened new opportunities for analyzing multi-modal data at the semantic level through natural language interactions. For example, GPT-4~\cite{achiam2023gpt} demonstrates the ability to reason over text, images, and JSON data within a unified conversational interface.
However, despite these advancements, current LLM-based approaches have yet to fully unlock the potential of multi-modal data analytics, primarily due to the following limitations:

%Recently, the rise of Large Language Models (LLMs) has made it possible to analyze the multi-modal data at the semantic level by interacting with human using Natural Language (NL). For instance, GPT-4~\cite{achiam2023gpt} can explore the text, image, and JSON data altogether by chating with human. However, existing LLM-based approaches have yet fully unlocked the potential to comprehend the multi-modal data due to the following limitations:

\noindent \textbf{Limitation 1: Limited Query Expressiveness.}
Existing approaches typically support querying only a subset of data modalities and suffer from limited expressiveness in capturing complex user intents. For example, ELEET~\cite{urban2024eleet} supports analytics over text and tabular data; Palimpzest~\cite{liu2025palimpzest} is restricted to text and image processing; and AOP~\cite{wang2025aop} primarily targets document and text data.
Underlying these systems are three main strategies for query translation: (1) mapping natural language (NL) directly to modality-specific operations~\cite{wang2025aop}, (2) translating NL to SQL using NL2SQL techniques~\cite{liu2024survey, fan2024combining}, or (3) requiring users to write declarative SQL-like query languages~\cite{liu2025palimpzest}.
However, these methods remain inadequate for data lakes, where user queries are often ambiguous and span a wide range of data modalities.

%\noindent \textbf{Limitation 1: Query Expressiveness.}  Existing approaches focus on querying data of a subset of modality and their capacity of query expressiveness is low.  For instance, ELEET~\cite{urban2024eleet} supports to analyze text and tables; Palimpzest~\cite{liu2025palimpzest} can just analyze text and images; AOP~\cite{wang2025aop} only targets at processing documents and text. Under the hood, they implement the query translation in three ways: (1) mapping the NL to the operations on specific modality~\cite{wang2025aop}, or (2) leveraging NL2SQL~\cite{liu2024survey, fan2024combining} technique to translate the NL to SQL queries, or (3) demanding users to compose the declartive SQL-like query languages~\cite{liu2025palimpzest}. Unfortunately, they are far from perfect because the query intent in data lake may have ambiguity and user queries can touch a wide specturm of modality .

\noindent \textbf{Limitation 2: High Inference Overhead.}
Most existing approaches rely on a single, unified LLM to process heterogeneous data modalities, which leads to significant inference overhead. This issue is attributed to the computational cost of state-of-the-art LLMs, such as GPT-4, which contains over a trillion parameters. Moreover, queries that span multiple modalities introduce complexity in query planning and execution, further impacting efficiency. From a query processing perspective, users typically expect low-latency responses, making it imperative to reduce inference overhead.

%\noindent \textbf{Limitation 2: Inference Overhead.} Existing approaches leverage one unified LLM to analyze the different kinds of data, leading to high inference overhead. This is mainly because advanced LLM models such as GPT-4 have more than a trillion parameters, and they are computationally expensive. Moreover, complex queries involving different modality make the query planning rather hard.  From the perspective of query processing, users normally demand high efficiency, thus reducing the inference overhead is critical.

% deep think techniques with reinforcement learning can further amplify the inference overhead

%\noindent \textbf{Limitation 3:  Data Freshness.} 
\noindent \textbf{Limitation 3: Knowledge and Data Staleness}
Data in the lake is often incomplete or becomes outdated over time, which undermines the freshness and reliability of analytical results. Moreover, the knowledge embedded in LLMs can also become stale as new data and emerging patterns (e.g., novel diseases or evolving user behavior) are not reflected in the original training data. For example, an LLM may fail to recognize symptoms of a newly discovered disease or interpret related medical images correctly. Existing approaches largely ignore this issue and operate under static assumptions. Therefore, it is essential to develop mechanisms for augmenting stale data and refreshing LLM knowledge to ensure up-to-date and contextually relevant analytics.

%The user data in the data lake may be incomplete and obsolete over time, leading to low data freshness and outdated data analytics. Moreover, the LLM knowledge can also be obsolete as new data is produced. For instance, if a disease with new symptoms emerges, it will never capture the features from corresponding images. Existing approaches overlook such an issue and solely studied the static cases. Hence, it is crucial to augment the data and refresh the LLM knowledge.

Model Context Protocol (MCP)~\cite{hou2025model} is a novel framework that standardizes the interaction among AI agents (e.g., knowledge-grounded LLMs), external tools (e.g., database engines, function calls), and data resources (e.g., structured and unstructured sources). By defining a unified interface, MCP enables seamless integration of real-time inputs, environmental variables, and domain-specific constraints, thereby supporting robust and scalable decision-making across diverse applications.
In the context of multi-modal data analytics, MCP provides an effective abstraction layer over heterogeneous data lakes. Users can interact with AI agents through natural language, while the analytics workload is offloaded to dedicated MCP servers, each tailored to handle a specific data modality. Moreover, with the growing ecosystem of MCP servers, thousands of off-the-shelf components will be readily available for processing multi-modal data efficiently.

%Model Contex Protocol (MCP)~\cite{hou2025model} is a novel framework designed to standardize and enhance the interaction among AI agents (e.g., knowledged LLM), extenal tools (e.g., database engine, function call), and resources (e.g., data sources). By establishing a unified interface, MCP enables users to adaptively integrate real-time inputs, environmental variables, and domain-specific constraints, ensuring robust and scalable decision-making across diverse applications. Concerning multi-modal data analytics, MCP offers a huge opportunity to provide an abstraction against multi-modal data lake where users can interact with AI agents and offloads the data analytics to MCP servers, and each MCP server can be responsible for one kind of modality. Moreover, as more and more MCP servers have been released, it provides thousands of off-the-shelf components to process multi-modal data.

% thereby enhancing the accuracy and efficiency while delivering high scalability..

%Rooted in principles of interoperability and modularity, the protocol addresses critical challenges in fields such as artificial intelligence, IoT ecosystems, and distributed computing, where seamless context integration is essential for accuracy and efficiency. MCP’s architecture prioritizes flexibility, allowing developers to tailor context parameters while maintaining compatibility with heterogeneous systems. As a result, it bridges the gap between static model assumptions and the fluidity of real-world scenarios, positioning itself as a foundational tool for next-generation adaptive technologies.

\begin{figure}[t!]
	\centering
	\includegraphics[width=0.85\linewidth]{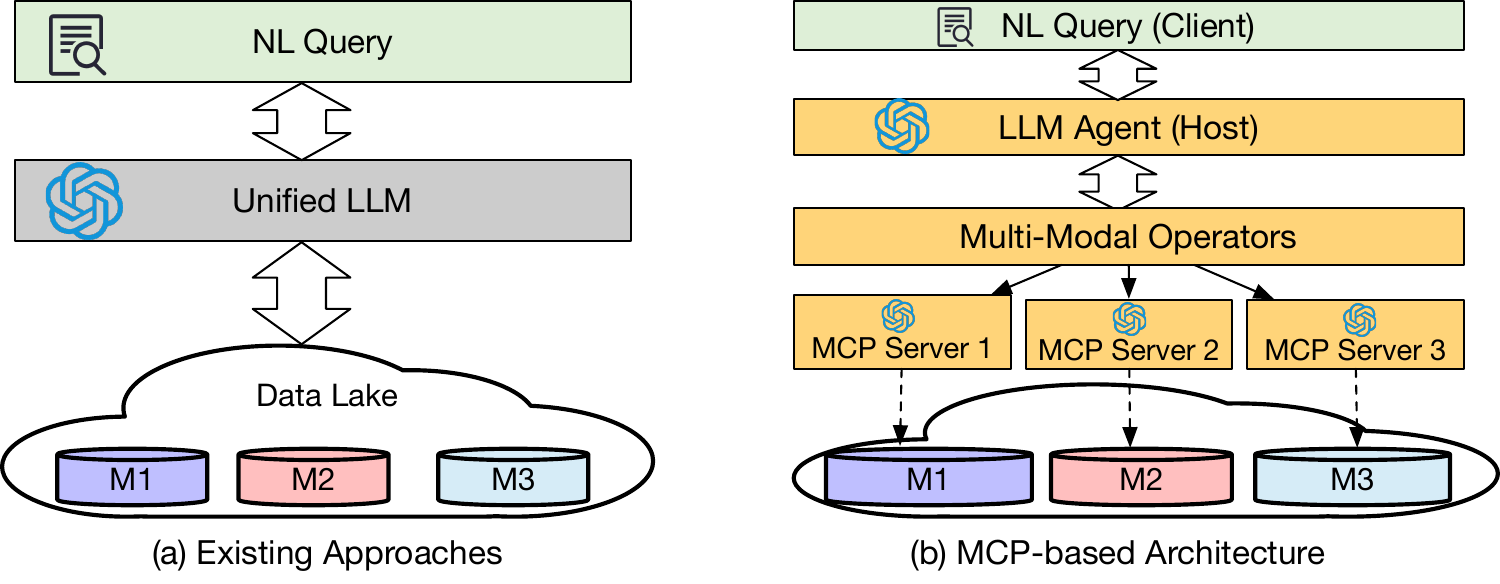}
	\caption{An Illustrate Example on MCP-based Multi-Modal Data Analytics.}
	\label{fig:motivation}
\end{figure}

Building upon the capabilities of MCP, we propose a novel multi-modal data analytics system, named \ours, to address the aforementioned limitations. 
\ours introduces a new architecture that leverages MCP to offload modality-specific analytical tasks to dedicated MCP servers.
As illustrated in Figure~\ref{fig:motivation}, \ours differs fundamentally from conventional approaches that rely on a single, unified LLM to process all modalities (e.g., M1, M2, M3 represent different data modalities). Instead, \ours adopts a client–host architecture: the client receives an NL query, and a host-side LLM agent interprets the user intent. This agent decomposes the query into a set of modality-specific operators and formulates a structured query plan. Each sub-plan is dispatched to an appropriate MCP server, where a tailored LLM, which is optimized for the corresponding modality, executes the analysis.
This modular and distributed design yields several key advantages: (1) higher inference accuracy by leveraging specialized models, (2) improved scalability through parallelism across servers, and (3) significantly reduced inference overhead compared to monolithic LLM-based solutions.

%In this paper, we envision a new multi-modal data analytical system, named \ours, to overcome the above-mentioned limitations. Particularly, we design a new architecture based on Model Context Protocol (MCP) that offloads the multi-modal operators to specific MCP server. Figure~\ref{fig:motivation} depicts the difference between existing approaches and MCP-based architecture. Compared with the existing approaches that leverage a unified LLM to analyze the multi-model data  with the NL query (Note that M1, M2, and M3 refer to data of different modality), MCP-based architecture follows a client-and-host architecture where client accepts the NL query and host utilizes a LLM agent to understand the query intent. Then, LLM agent orchestrates the multi-modal operators to a query plan, and each subplan is send to an MCP server that leverages a tailored LLM to process the data of specific modality, delivering higher accuracy, higher scalability, and lower inference overhead.  

%However, there exist three challenges as follows:

In summary, this paper has the following key contributions:

(1) \textbf{MCP-Based Multi-Modal Analytical System.} 
We propose a novel multi-modal data analytics framework built upon the Model Context Protocol (MCP), which addresses the limitations of existing LLM-based systems. The core idea is to assign each MCP server a tailored LLM optimized for a specific data modality, embracing the principle that ``one size does not fit all''.

%(1) We propose a novel MCP-based multi-modal data analytical system that can overcome the limitations of existing LLM-empowered approaches. The key idea is to let each MCP server leverages a tailored LLM to handle the data of specific modality because of  the principle of ``One size does not fit all".

(2) \textbf{Semantic Operator Hierarchy.}
We introduce a hierarchical set of semantic operators that spans structured, semi-structured, and unstructured data. This design enables \ours to support advanced tasks such as cross-modal joins, while maintaining compatibility with existing operators for relational data, documents, graphs, images, audio, and video, thus avoiding ``reinventing the wheels''.

%(2) We introduce a hierarchy of a set of semantic operators, aiming to cover a wide range of data modality, across structured, semi-structured, and unstructured data. The operator hierarchy not only enables \ours to process new operators such as cross-modal join, but also keep the existing operators usable, such as operators on relational, document, graph, images, audio, video, thereby avoiding ``reinventing the wheels". 

(3) \textbf{Query-Driven Model Optimization and Data Discovery.} We develop a query-driven fine-tuning strategy to optimize the reasoning ability of LLMs on each MCP server. Moreover, we propose a unified embedding-based semantic representation combined with a hybrid indexing mechanism for multi-modal data discovery.

%(3) We design a query-driven fine-tuning method to enhance the LLM capability of each MCP server. Moreover, we leverage an embedding method to unify the semantic representation and then we build a hybrid index to speed up the process of data discovery.

(4) \textbf{Dynamic Knowledge and Data Refreshing.} We propose a twofold updating mechanism to ensure freshness of both the data and the LLMs. First, we leverage deep research techniques to enrich the data lake with the latest documents and web content. Second, we introduce a lightweight editing mechanism (supporting insert, update, and delete operations) augmented by machine unlearning techniques to refresh LLM knowledge.

%(4) We propose a updating mechanism to refresh the data and LLM knowledge. First, we harness the deep research technique to enrich the data lake with newest documents and web pages. Second, we propose to edit the LLM knowledges with the support of the insert, update, and delete. Particularly, we utilize machine unlearning technique to refresh the LLM knowledges.

% Specifically, we first define a hierarchy of semantic operators upon multi-modal data and develop an NL2Operator translator using AI agent. We then propose an MCP-based operator execution architecture where each MCP server can host tailored foundation models for data of specific modality, thereby enhancing the accuracy and efficiency while delivering high scalability. Next, we propose a updating mechanism by harnessing the deep research and machine unlearning  techniques to refresh the data lakes and LLM knowledges, with the goal of balancing the data freshness and inference efficiency.

%\begin{example}
%	Consider an example
%\end{example}

% To address aforementioned challenges, we propose \ours, a novel MCP-based multi-modal data analytics system in data lakes.

% 2.  MCP Motivation

% 3. Challenges and Solutions

\section{Preliminaries}

\subsection{Problem Definition}

\textbf{Multi-Modal Query Processing (MMQP).}
Given a natural language (NL) query $Q$ over a collection of datasets $\mathbb{D}$, where each dataset $D_i \in \mathbb{D}$ is associated with a modality $M_i \in \mathbb{M}$, the goal is to compute a result set $R$ that satisfies the query predicates, where each result tuple $r \in R$ may contain data items spanning multiple modalities in $\mathbb{M}$.
To evaluate the performance of MMQP, we adopt three standard metrics: (1) recall, which measures the completeness of the retrieved results, (2) precision, which quantifies the ratio of the correct ones among the returned results. and (3) latency, which reflects the query execution efficiency.

%\textbf{The Problem of Multi-Modal Query Processing (MMQP).} Given a NL query $Q$ against a set of datasets $\mathbb{D}$ where each $D_i \in \mathbb{D}$ is associated with a modality $M_i\in \mathbb{M} $, the goal is to find the multi-modal results $R$ satisfying the predicates of original query and service requirements such as minimum response time. Note that each tuple  $r \in R$ contains data from $\mathbb{M} $. We adopt the common metrics: query recall and latency, in order to quantify the query processing's accuracy and efficiency, respectively.

%This paper considers a structured table T with n attributes (i.e., columns), denoted by fA1; A2; : : : ; Ang, and m tuples (i.e., rows), denoted by ft1; t2; : : : ; tmg. In particular, we denote the value of the j-th attribute in tuple ti as tij. Moreover, we consider a textual hypothesis (or claim), denoted by C. The problem of table-based fact verification is, given a pair (C; T ) of claim C and table T, to determine whether table T can be used as evidence to support or refute claim C. Note that claim C may be either as simple as describing one tuple in table T, or as complex as aggregating or comparing multiple tuples in T. 

In general, the MMQP problem has the following challenges:

\noindent \textbf{Challenge 1: Cross-Modality Query Planning.}
Unlike traditional query planning in relational databases that focuses solely on relational operators, orchestrating operator pipelines across multiple data modalities is more complex. This complexity arises from the heterogeneous nature of operators and data formats. Therefore, MMQP requires an effective yet lightweight mechanism to generate reasonable execution plans. 
%To tackle this, we propose a sampling-based query planning approach that balances planning latency and execution quality.

%\noindent \textbf{Challenge 1.} Unlike traditional query planning in relational databases that solely considers relational operators, orchestrating operator pipeline in the context of multiple modality is much more complex. Therefore, it calls for an effective yet lightweight way to generate a reasonable execution pipeline for MMQP. To handle such a challenge, we propose a sampling-based query planning to balance the planning time and plan quality.

\noindent \textbf{Challenge 2: Efficient Inference on Large-Scale Data.}
Compared with using a general-purpose LLM with a massive number of parameters, employing a tailored, relatively lightweight LLM for modality-specific data can significantly reduce inference overhead. However, this approach still encounters scalability challenges when dealing with large volumes of data, particularly unstructured data such as text, images, or videos.

%\noindent \textbf{Challenge 2.} Compared with using a general LLM with enormous parameters, leveraging a tailored relatively-smaller LLM to analyze the specific data of modality can indeed reduce the inference overhead. However, it still face a challenge when it comes to a large volume of data, especially for unstructured data. To address such a challenge, we propose an embedding index that contains both semantic and metadata information, allowing for pruning the unrelated data before the data analytics.

\noindent \textbf{Challenge 3: Data and Knowlege Freshness.}
High-quality data in the data lake is often insufficient to precisely and comprehensively answer user queries. Therefore, it is crucial to continuously update both the data lake and the LLM knowledge within each MCP server. However, open-domain data sources are vast and challenging to explore, making integration into the data lake cumbersome and resource-intensive.

%Data of high quality in the data lake is often insufficient to answer the user query precisely and comprehensively. Thus, it is important to update the data lake and LLM knowledge concerning each MCP server. Unfortunately, the data source in the open-domain is rather huge to explore and is cubmersome to integrate into the data lake. To address such a challenge, we resort to augment the data using deep research technique and then propose a query-driven fine-tuning method to update the importance of the corresponding LLM parameters. We also utilize the machine unlearning technique to erase the dirty data in the LLM.

\subsection{Related Work}

In recent years, LLMs have made remarkable advances in natural language understanding, generation, and reasoning, opening up new opportunities for multi-modal data management. By expressing their query requirements in natural language, users can interact with systems that leverage LLMs to interpret query intent and dynamically orchestrate external databases, knowledge bases, or APIs to execute complex analytical tasks.

%In recent years, large language models (LLMs) have achieved groundbreaking progress in natural language understanding, generation, and reasoning, bringing new opportunities for the development of multi-modal data management. Users describe their query requirements in natural language, and the system leverages LLMs to interpret the query intent while dynamically coordinating external databases, knowledge bases, or APIs to execute complex query tasks.

% Thanks to the powerful inference capabilities of  LLMs, an increasing number of research efforts have begun exploring the use of LLMs as core components to drive the querying and analysis of multi-modal data, giving rise to a series of LLM-centric multi-modal query systems. Such systems typically employ natural language as the primary query interface.

CAESURA~\cite{urban2023caesura} (also known as ELEET~\cite{urban2024eleet}) utilizes LLMs as multimodal query planners that translate NL inputs into executable multimodal query plans. Its key idea is leveraging LLMs to understand user intent and dynamically construct efficient query workflows tailored to diverse data modalities. Building upon this, Gorilla~\cite{patil2024gorilla} enhances LLMs’ reasoning and tool-use capabilities by incorporating retrieval-augmented generation (RAG), enabling real-time access to external knowledge bases and APIs during multimodal query processing.
Toolformer~\cite{schick2023toolformer} further advances LLM tool integration by fine-tuning models with limited examples, empowering them to autonomously invoke external APIs, such as database engines and computational tools, during generation. This overcomes LLMs' traditional limitation to static text generation, enabling dynamic interaction with external data sources for complex analytical tasks.
ThalamusDB~\cite{jo2024thalamusdb} introduces a hybrid architecture combining a dedicated query planner with deep learning inference. It allows SQL queries to include NL predicates, offloading parts of the execution to neural models for image and text processing, while leveraging approximate query processing (AQP) to balance efficiency and accuracy. However, it still relies on manual annotations, which may hinder scalability.
PALIMPZEST~\cite{liu2024declarative, liu2025palimpzest} extends declarative query languages to express AI workloads, generating optimized plans that jointly consider the cost of both AI inference and relational operations. AOP~~\cite{wang2025aop, wangidatalake} proposes defining semantic operators over documents, text, and structured tables, and employs LLMs to iteratively generate executable query plans for multi-modal content.

However, existing works  have three main limitations:
First, they support only a few data modalities and are hard to extend, i.e., adding new modalities or operators often requires heavy manual work. In contrast, \ours supports a wide range of multimodal operators with a scalable MCP-based architecture.
Second, most systems use a single LLM to handle all data types, which leads to low efficiency and accuracy. \ours instead uses tailored LLMs for different modalities, improving performance.
Third, they focus on static data and ignore updates. \ours addresses this by automatically enriching the data and refreshing model knowledge over time.

%To conclude, existing works has three main limitations. First, they support only a few modality and fail to provide a comprehensive and efficient way to analyze multi-modal data. That is, users must manually modify many lines of code to support new modality or new operator. In contrast, \ours supports a wide spectrum of multi-modal analytical operator and provide a scalable and disaggregated architecture based on MCP. Second, they mainly utilize one unified LLM to handle multi-modal query processing, leading to inefficient ad inaccurate query processing. On contrary. \ours develops the tailored LLM model to handle specific data of modality. Third, they barely focus on the static cases while neglecting the data changes and knowledge updating. While \ours enables to enrich the data and update the model automatically.

\section{\ours's Overview}
\begin{figure}[t!]
	\centering
	\includegraphics[width=0.85\linewidth]{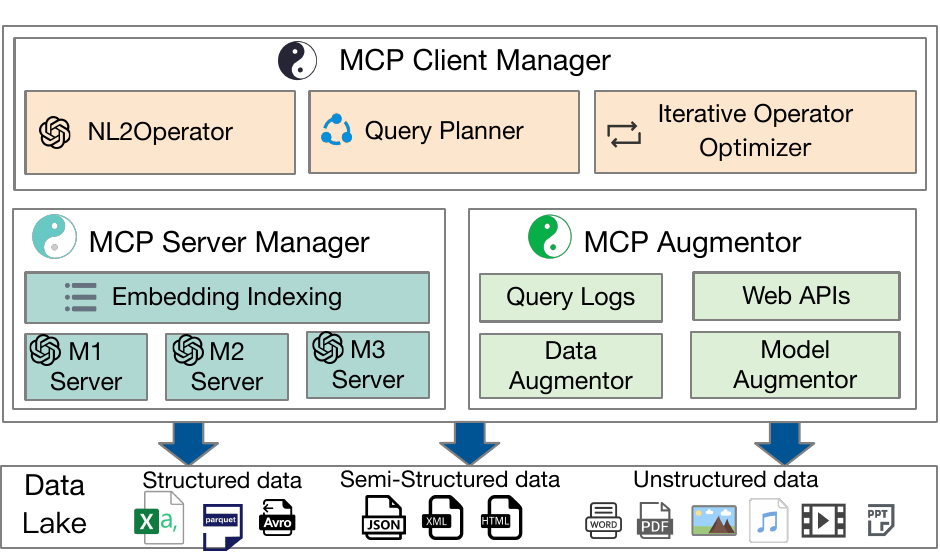}
	\caption{An Overview of \ours.}
	\label{fig:overview}
\end{figure}

% \fanj{we need a running example to demonstrate the effectiveness of MCP.}

Figure~\ref{fig:overview} presents an overview of \ours, which consists of three main components: \textbf{MCP Client Manager}, \textbf{MCP Server Manager} and \textbf{MCP Augmentor}. In the following, we introduce them in details.

\subsection{MCP Client Manager}  

\subsubsection{NL2Operator} 
Understanding user intent is critical for accurate query answering. However, neither SQL-like declarative languages nor UDF-based procedural languages are well-suited for this task. This is due to two main challenges:
First, NL queries are inherently ambiguous and difficult to interpret accurately.
Second, translating NL queries into complex declarative or procedural languages increases the likelihood of inaccuracies or errors.

To address this, we propose NL2Operator, a method that defines a hierarchical set of semantic operators, each linked to specific MCP servers. An LLM-based agent maps user queries to these operators based on query intent. For example, semi-structured data processing is handled by an intermediate MCP server, which can either process the query directly or delegate it to specialized sub-servers (e.g., for JSON, HTML, XML).
This hierarchical design offers three advantages:
(1) Scalability and flexibility, by modularizing the processing across different servers;
(2) High concurrency support, through distributed workload balancing;
(3) Simplified translation, by mapping NL to high-level modality-specific operators, reducing the burden on the LLM.

%Comprehending users' querying intent is the key to answer the query precisely. Unfortunately, neither SQL-like query language nor the UDF-based procedual language can well handle such a problem. The reason is two-fold. First, the NL-based user query has ambiguity, and it is hard to accurately understand the actual query intent concerning the various domains such medical, game, finance. Second, translating the NL query to another query language may increase the probability of translation error due to the complexity of the declarative/procedual query language, thus it is challenging to generate valid queries with high probability.  To this end, we propose a NL2Operator approach, which first defines a hierarchy of semantic operators binding to different MCP servers, then leverage LLM agent to map the query intent to the MCP servers. For instance, we bind the semi-structured data processing to an intemediate MCP server which can either process the data itself or offload the query processing to its children MCP servers which includes a JSON-oriented server, an HTML server, and an XML server.  Such an organization brings not only scalability and  flexibility, but also can handle high concurrent query requests with proper workload balance. Furthermore, it eases the burden of LLMs as the translation is simplifed to map the NL language to a high-level modality. In addition to semi-structured data, we create the hierarchy for structured data and unstructured data as well.

\subsubsection{Query Planner}  

The goal of query planner is to find the execution plan with the lowest cost, thereby improving overall efficiency. Based on a hierarchy of semantic operators, we construct a directed acyclic graph (DAG) to represent the query workflow. We then implement a sampling-based cost estimation optimizer, which includes selectivity estimation, cost modeling, plan evaluation, and final plan selection.

The optimization process works as follows: First, an operator cost sampler collects runtime statistics, such as per-tuple processing time and selectivity, for each operator. Using this data, the optimizer builds a latency-based cost model to evaluate candidate plans. The plan with the lowest estimated execution time is selected for execution. This sampling-based approach supports accurate, data-driven cost prediction and adaptive plan selection. It offers two key advantages: (1) it maintains reliable cost estimates under different data distributions and runtime conditions, and (2) it produces more deterministic plans compared to LLM-based query planning.

%The goal of query optimization is to identify the query plan with the lowest execution cost, thereby improving query efficiency. Given a set of hierarchy operators, we build a DAG (directed acyclic graph) as the basic workflow, then we designed and implemented a sampling-based cost estimation query optimizer, whose core components include selectivity estimation, cost modeling, query plan evaluation, and optimal plan selection. 
%
%The overall query optimization workflow is as follows: First, the query optimizer dynamically collects execution costs for each query operator (including per-tuple processing time and selectivity) through an operator cost sampler. Next, the optimizer constructs a latency-based cost model using the sampled data and evaluates each candidate query plan against this model. Finally, the optimizer selects the plan with the minimal estimated execution time as the final execution strategy. This approach enables data-driven cost prediction and adaptive plan selection, balancing optimization overhead with execution performance gains. The sampling mechanism has two merits. First, it ensures cost estimates remain accurate under varying data distributions and runtime conditions. Second, it has more determinstic output than LLM-based query planning.

\subsubsection{Iterative Operator Optimizer}  

\begin{figure}[t!]
	\centering
	\includegraphics[width=0.85\linewidth]{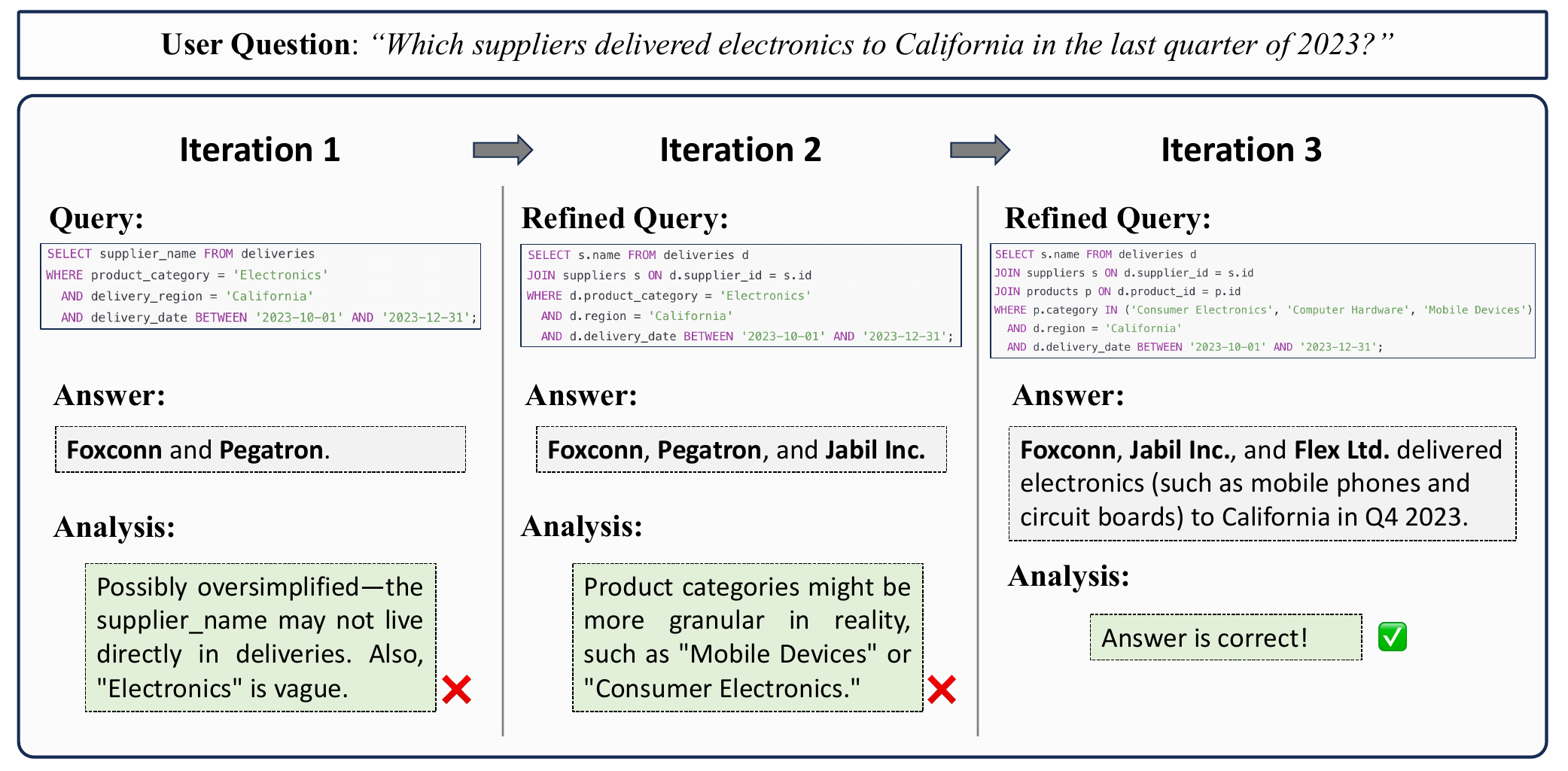}
	\caption{An illustrative example of iterative operator optimization in \ours.}
	\vspace{-1em}
	\label{fig:iterative_rag}
\end{figure}

Each MCP server is responsible for processing data from a specific modality, autonomously interacting with its underlying data source, whether structured (e.g., SQL-based) or unstructured (e.g., document corpus), via an Iterative operator optimization \cite{NEURIPS20206b493230,shuster2022language}. 
Unlike traditional pipelines that rely on a single retrieval pass, \ours introduces a bidirectional communication channel between the MCP server and the MCP client (or host). This feedback loop enables the server to request prompt refinements or clarifications from the upstream controller when retrieval results are insufficient, supporting a dynamic, feedback-driven, and iterative optimization process.

%Each MCP server is dedicated to processing data from a specific modality by autonomously interfacing with its associated structured (e.g., SQL-based) or unstructured (e.g., document corpus) database through an Iterative Retrieval-Augmented Generation (RAG) loop \cite{NEURIPS20206b493230,shuster2022language}. Unlike conventional RAG pipelines that perform retrieval in a single pass, our architecture incorporates a bidirectional communication mechanism between the MCP server and the MCP client/host. This mechanism allows the server to request prompt refinements or clarifications from the upstream controller when retrieval results are inadequate—thereby enabling a feedback-driven iterative RAG process.

An example of the iterative architecture in \ours is illustrated in Figure~\ref{fig:iterative_rag}. The process starts when the MCP server receives a sub-task from the central LLM agent. Guided by the task’s semantics and structure, the server formulates an initial query and retrieves candidate results from its local database. These results are evaluated along multiple dimensions, including coverage, redundancy, ambiguity, and informativeness. If deficiencies are detected (e.g., sparse or misaligned results), the server engages a reasoning-based refinement loop to revise the query and improve alignment with the task’s underlying intent. This iterative cycle continues until the result set meets a predefined confidence threshold. Moreover, we employ a curriculum learning strategy that gradually increases task complexity, from simple key-value lookups to multi-hop reasoning over relational and multimodal data, thus guiding the agent through progressively more abstract levels of query planning and evaluation.

\subsection{MCP Server Manager}

\subsubsection{Embedding Indexing}  
For high-dimensional vector data, traditional vector indexing methods~\cite{DBLP:journals/pacmmod/AziziEP25} typically construct a graph structure by linking neighboring data points based on nearest-neighbor distance. However, when metadata-based filters are applied, many of these connections become invalid, resulting in disconnected subgraphs. This leads to search failures, as the index may no longer be able to reach the true nearest neighbors. To address this, \ours introduces a filter-aware vector indexing approach designed to maintain search effectiveness even under filtering constraints. The key idea is to selectively augment the vector graph by adding edges that connect only those data points that comply with the filtering conditions. This ensures that the search process remains efficient while achieving high recall. Figure~\ref{fig:iterative_rag} illustrates the core concept. The key components of our method include:

%For high-dimensional vector data, traditional vector indexing connects all neighboring data points through nearest-neighbor distance calculations in a graph structure. However, when metadata filtering conditions are applied, many data points become disconnected, causing the vector index to fail and preventing the search from traversing to the true nearest neighbors. \ours aims to develop a filter-aware vector indexing approach that addresses search failure under filtering conditions. The proposed method augments the vector graph index by selectively adding compliant connections that satisfy the filtering constraints, thereby maintaining search efficiency while preserving reasonable query recall rates. Figure~\ref{fig:iterative_rag} gives an illustration of our key ideas, and the key aspects of our approach include:

(1) Condition-aware graph augmentation: Dynamically reinforce valid connections among filter-compliant nodes to ensure index traversability even after metadata-based filtering is applied.

(2) Hybrid search robustness: Maintain a balance between filtering precision and vector search accuracy by preserving essential traversal paths within the index.

(3) Recall preservation: Ensure that the augmented index structure delivers competitive nearest-neighbor retrieval performance, comparable to that of unfiltered searches.

This approach extends traditional vector search to support constrained queries where both semantic similarity (via embeddings) and structured criteria (via metadata) must be jointly satisfied.

\begin{figure}[t!]
\centering
\includegraphics[width=0.8\textwidth]{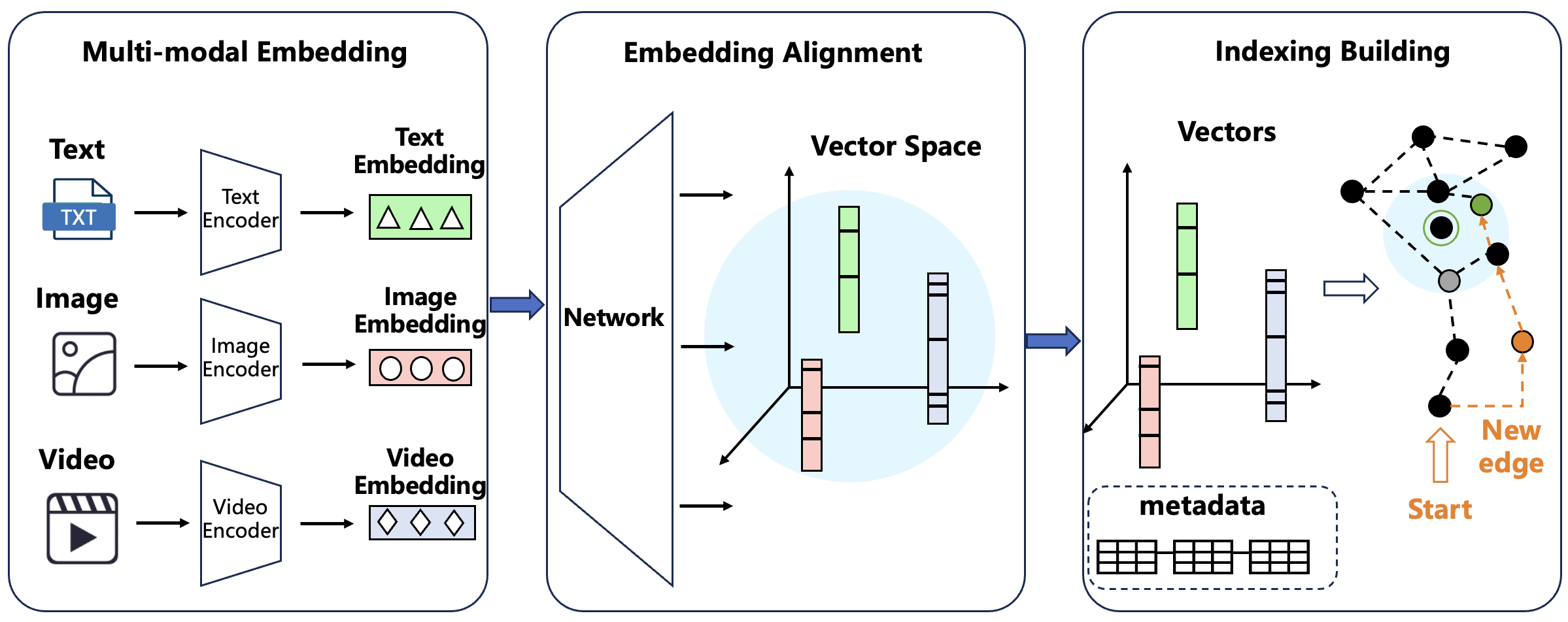}
\caption{Embedding of Multi-Modal Data.}
\vspace{-1em}
\label{fig:iterative_rag}
\end{figure}
%
%\textbf{(1) Condition-aware graph augmentation: } Dynamically reinforcing valid connections between qualified nodes to ensure traversability post-filtering.
%
%\textbf{(2) Hybrid search robustness: } Balancing the trade-off between filtering strictness and vector search accuracy by intelligently preserving critical pathways.
%
%\textbf{(3) Recall preservation:} Algorithmically guaranteeing that the modified index structure maintains competitive nearest-neighbor retrieval performance compared to unfiltered scenarios.
%
%This methodology extends conventional vector search capabilities to constrained query scenarios where both semantic similarity (via vectors) and structured conditions (via metadata) must be jointly optimized.

\subsubsection{Multi-Modal MCP Servers}  
Multi-Modal MCP Servers deal with heterogeneous data ecosystems by unifying structured (e.g., relational databases), semi-structured (e.g., JSON logs, XML), and unstructured data (e.g., text, images, sensor streams) under a single adaptive framework. Leveraging LLMs and multi-modal AI architectures, these servers dynamically interpret, classify, and contextualize diverse data formats through techniques such as natural language processing (NLP) for unstructured text, computer vision for visual data, and graph-based reasoning for structured relationships. By embedding semantic understanding into the MCP layer, the system autonomously generates metadata, enforces cross-modal data governance policies, and enables federated queries that bridge tabular sales records, semi-structured IoT telemetry, and unstructured social media feeds. For instance, an LLM-powered MCP Server could correlate customer sentiment (extracted from unstructured reviews) with structured transactional data to optimize supply chain decisions, while simultaneously parsing semi-structured maintenance logs to predict equipment failures.

\subsection{MCP Augmentor}  
\subsubsection{Data Augmentor}

% After executing queries on modality-specific databases, each MCP server performs an autonomous update procedure that includes both data augmentation and model capability enhancement, as shown in Figure \ref{fig:update}. In the data augmentation phase, to ensure the data lake remains fresh and semantically rich, \ours adopts a query-driven augmentation method that conducts deep research  following query execution. During this phase, the server leverages web crawlers and API-based search agents to retrieve relevant content from dynamic sources such as academic papers, technical blogs, and authoritative web pages. Retrieved documents such as research papers, official documentation, and high-quality blogs are parsed and deduplicated. Named entity recognition (NER) and semantic hashing are used to identify novel entities or facts not yet present in the data lake. Only validated and novel knowledge entries are added to the data lake, indexed by modality and timestamp to support time-sensitive reasoning.

\begin{figure}[t!]
	\centering
	\includegraphics[width=0.7\linewidth]{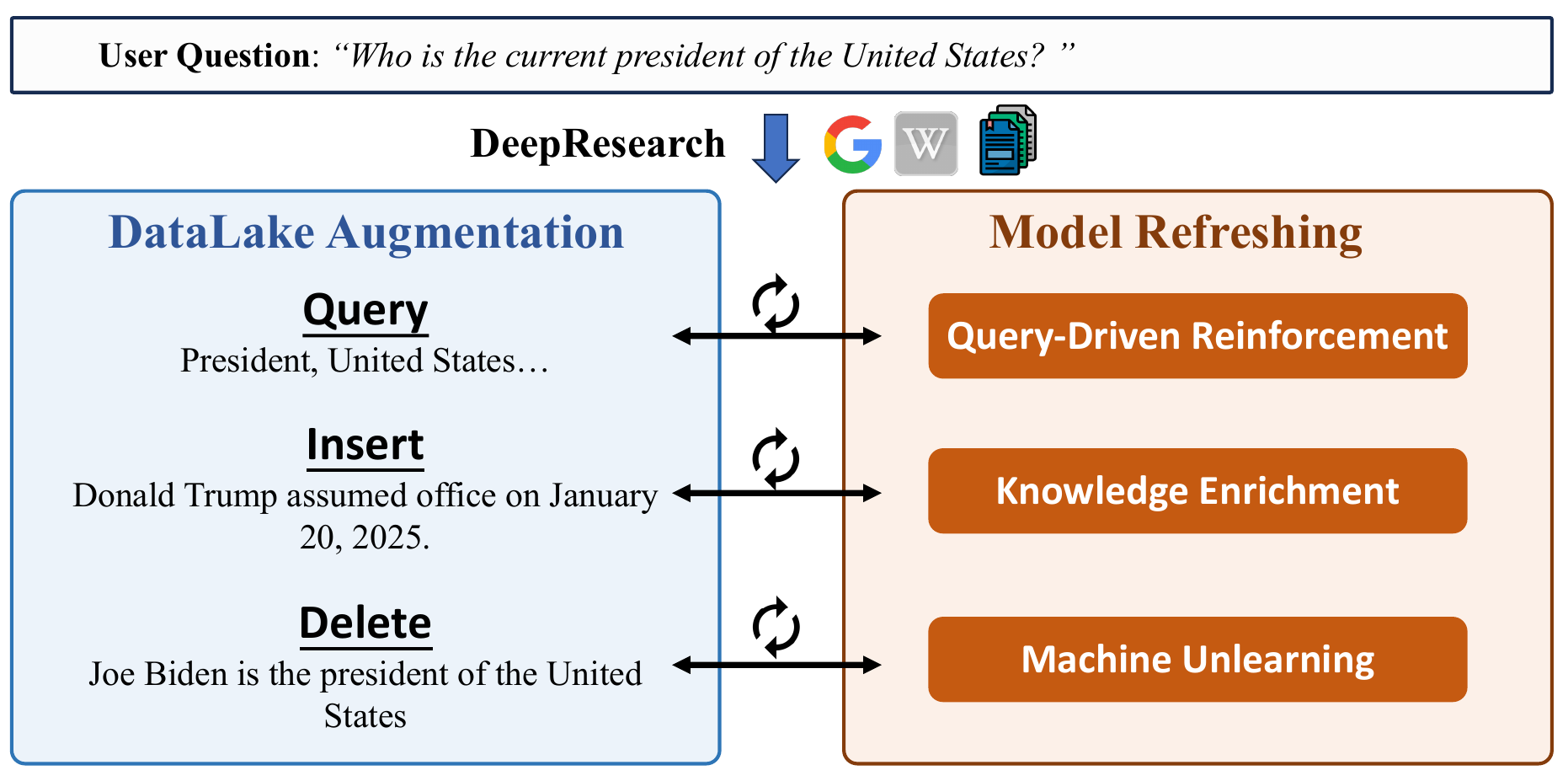}
	\caption{Data Augmentation and Model Refreshing in \ours.}
	\label{fig:update}
\end{figure}

After executing queries on modality-specific databases, each MCP server autonomously initiates an update procedure involving both data augmentation and model capability enhancement, as shown in Figure~\ref{fig:update}. In the data augmentation phase, \ours adopts a query-driven retrieval-then-synthesis strategy to keep the data lake semantically enriched and aligned with the latest developments. Upon completion of a user query, the MCP server triggers targeted information harvesting routines. These leverage both web crawlers and structured API-based agents configured to access dynamic and authoritative sources such as arXiv papers, GitHub repositories, Stack Overflow, and enterprise technical documentation.

%After executing queries on modality-specific databases, each MCP server autonomously performs an update procedure encompassing both data augmentation and model capability enhancement, as illustrated in Figure \ref{fig:update}. In the data augmentation phase, \ours employs a query-driven retrieval-then-synthesis strategy to ensure that the data lake remains semantically enriched and up-to-date with emerging information. Upon completing a user query, the MCP server launches targeted information harvesting routines that leverage both web crawlers and structured API-based agents. These agents are configured to access dynamic and authoritative sources, including arXiv paper, GitHub repositories, Stack Overflow, and corporate technical documentation.

The collected documents undergo a multi-stage processing pipeline designed to transform unstructured content into validated, semantically indexed knowledge.

%The collected documents are subjected to a multi-stage processing pipeline that transforms unstructured sources into validated, semantically indexed knowledge. 

\textbf{(1) Structural Reconstruction:}
The pipeline begins with document structure parsing, utilizing tools such as ScienceParse \cite{scienceparse} and GROBID \cite{grobid} to extract key structured elements, including titles, abstracts, section headers, equations, tables, figures, and code snippets. These tools employ a combination of rule-based heuristics and machine learning techniques to reconstruct the hierarchical organization of scientific and technical documents from raw formats such as PDFs and HTML.

% \fanj{why scientific and technical documents are relevant to \ours???}

%The pipeline begins with document structure parsing, where tools such as ScienceParse \cite{scienceparse} and GROBID \cite{grobid} are used to extract structured elements including titles, abstracts, section headers, equations, tables, figures, and code snippets. These tools leverage rule-based and machine learning approaches to reconstruct the hierarchical layout of scientific and technical documents from raw PDFs or HTML content.

\textbf{(2) Redundancy Elimination:}
In the redundancy elimination stage, a hybrid strategy combines MinHash-based fingerprinting for fast approximate set similarity detection with dense embedding-based retrieval. Specifically, documents are first fingerprinted using MinHash signatures generated over token-level n-grams. In parallel, dense vector embeddings are computed using models like Sentence-BERT \cite{reimers2019sentencebert}, and clustered using HNSW-based approximate nearest neighbor search (via libraries like FAISS) to detect semantically similar passages. Duplicate or near-duplicate entries are then filtered out based on a configurable similarity threshold.

\textbf{(3) Entity Extraction:}
Next, domain-adapted named entity recognition (NER) is applied using fine-tuned transformer-based models trained on labeled datasets from scientific and software domains. To resolve synonyms and disambiguate entities across sources, we apply context-aware entity linking using string similarity, co-occurrence statistics, and external knowledge bases (e.g., DBpedia, PapersWithCode APIs). To identify novel concepts and facts, we compute concept-level hashes using SimHash, which captures the semantic footprint of passages. These hashes are used to cluster conceptually similar content and measure deviation from existing entries in the data lake. We also apply graph-based semantic clustering over the extracted entities and their relationships to reveal emergent topics not previously indexed.

\textbf{(4) Data Augmentation:}
In the final augmentation and indexing stage, each candidate entry is validated across multiple independent sources to ensure factual consistency and reliability. Only entries that are corroborated—i.e., referenced in at least two unrelated sources—are retained. The resulting validated knowledge units are indexed along multiple axes: modality (e.g., text, code, image), source credibility, and temporal metadata (e.g., crawl time, publication date). This indexing supports temporal knowledge reasoning, such as prioritizing more recent findings or applying decay-weighted relevance during downstream model retrieval.

\subsubsection{Model Refreshing}

In the Model Refreshing phase, to align model knowledge with the evolving data lake, we introduce a modality-specific parameter update mechanism tailored to three data operations: querying, insertion, and deletion. Thanks to the built-in subscription function of MCP server, the MCP client can monitor the changes of data resources via subcription, then we can conduct the model refreshing.

\textbf{(1) Query-Driven Reinforcement:} To enhance model performance based on frequently asked or emergent user intents identified from query logs, we first perform an in-depth analysis to extract the most common and critical intent patterns. These patterns are then leveraged to generate task-specific training samples. Furthermore, to ensure that these synthesized samples effectively contribute to the model's learning process, we apply importance sampling during the fine-tuning phase. Importance sampling assigns a higher weight to these specific training examples, based on their frequency and relevance in real-world use cases. This mechanism helps the model prioritize the learning of high-value user intents while preventing overfitting to less common or irrelevant patterns. By combining these methods, we significantly improve the MCP server’s ability to understand and respond to emergent user needs in real-time applications.

\textbf{(2) Insert via Knowledge Enrichment:} When new knowledge is ingested into the data lake, it undergoes a structured transformation where relevant text spans or structured tuples are converted into instruction-style training samples. These samples are then systematically appended to the fine-tuning dataset. The conversion process involves identifying key knowledge components and framing them as prompts and expected responses, ensuring that the newly introduced information is aligned with the model’s reasoning mechanisms. By incorporating these structured samples, the model is exposed to explicit tasks that require reasoning over the updated knowledge, enabling it to more effectively process and apply this new information during inference. This method directly enhances the model’s capacity for knowledge integration, thereby improving its performance on tasks that necessitate reasoning over both the pre-existing and newly added knowledge. Furthermore, this fine-tuning approach mitigates the risk of knowledge forgetting, as the model learns to consistently reference the most recent data alongside the foundational knowledge it was initially trained on.

\textbf{(3) Deletion via Machine Unlearning:} To address the removal of obsolete or sensitive knowledge, we employ advanced techniques such as gradient ascent unlearning or influence function-based data deletion to effectively erase the target information without compromising the integrity of the remaining model knowledge \cite{zhang2023recommendation}. Specifically, we first compute data influence scores, which quantify the contribution of individual training samples to the model's predictions. These scores are utilized to identify the most influential data points that need to be removed. Subsequently, we employ a fine-tuning strategy where negatively-weighted gradients are applied to the model, directing it to unlearn the targeted knowledge. This process is executed in a manner that is mindful of the need to preserve the model's retained knowledge, ensuring that the unlearning of specific information does not lead to catastrophic forgetting of other critical data or degrade the overall performance \cite{chourasia2023forget}. The integration of these techniques enables a controlled and efficient forgetting mechanism that minimizes interference with the model's pre-existing knowledge, making it an ideal approach for handling sensitive or outdated information.

This unified update mechanism ensures that each MCP server maintains alignment between its internal LLM and its evolving modality-specific data lake, enabling accurate, efficient, and up-to-date multi-modal response. In particular, it facilitates synchronized parameter tuning and schema-aware knowledge injection across modalities such as vision, audio, and text, thereby preventing semantic drift and enhancing the consistency of cross-modal representations. This design also supports incremental updates without requiring full model retraining, significantly reducing computational overhead while preserving reasoning fidelity.

\section{Preliminary Experiments}
In this section, we evaluate the performance of a prototype of \ours, demonstrating the efficiency and effectiveness of the proposed MCP-based architecture on handling multi-modal data.

\subsection{Experimental Setup}
\textbf{Dataset.} We use the Craigslist furniture dataset~\cite{jo2024thalamusdb}, which contains both relational and image data, to evaluate execution accuracy and latency of \ours. The dataset comprises 3000 listings from the ``furniture" category in website craigslist.org, detailing items such as sofas, tables, and chairs. It consists of two tables: \textit{furniture} and  \textit{image}. The \textit{furniture} table includes information of each entity; with each furniture has one or more images related to it, the \textit{image} table contains the path of images corresponding to each furniture. 

We design three queries to verify that \ours can demonstrate advantages concerning varied selectivity and task complexity. For simplicity, the query plan is fixed by performing a filter on the furniture table and then conducting matches on images with specified predicates. The intermediate result size refers to the size of set filtered by relational predicates, and image predicate refers to the image classification task. Table \ref{tab:sql} gives the summary of the workload. 

\textbf{MCP Client Setting and Server Setting.}
As introduced before, we implement a disaggregated MCP-based data analytical system. Specifically, we leverage the ChatGPT API~\cite{gptapi} in the MCP client, which is built upon the GPT-4.1 architecture~\cite{gpt4}. Additionally, We use the Qwen2.5-VL-7B~\cite{Qwen} as an MCP server to analyze complex images. We also deploy another MCP server~\cite{postgresqlmcp} to manage the structured data and semi-structured data based on a PostgreSQL database.

\textbf{Baseline.}
We compare our method to a pure GPT-4.1 model, which handles the query translation and data analysis altogether. On the contrary, our approach splits the task to three sub-tasks with MCP: query translation, table filtering, and image analysis. Note that as GPT-4.1 is not good at processing tabular data, we utilize PostgreSQL to help it handle the table filtering for a fair comparison. 

%to show that \ours achieves a much better performance with access to the powerful, local-deployed vision-language model.

\textbf{Evaluation Metric.}
For the evaluation of \ours, we employ three metrics: recall, accuracy and latency, which are respectively employed to evaluate correctness and efficiency. We compare the result set retrieved by client with the ground truth, to compute recall and precision. We also evaluate latency of performing the queries in an end-to-end fasion.

%calling the image-analyzer MCP server, as it is not only the bottleneck, but also where our approach makes the whole difference.

\begin{table}[]
\caption{Summary of the Workload }
\centering
\begin{tabular}{llll}
\hline
Query ID & Intermediate Size & Image Predicate & NL Query                           \\ \hline
1      & 33                        & images with two chairs                  & Find a set of two chairs             \\
2      & 126                      & images with a black chair            & Find a black leather chair         \\
3      & 347                      & images with table and chair        &  Find a set of wood table and chair \\ \hline
\end{tabular}
\label{tab:sql}
\end{table}

\subsection{Experimental Results}

\noindent
\begin{minipage}[t]{0.3\textwidth}
    \centering
    \includegraphics[width=\textwidth]{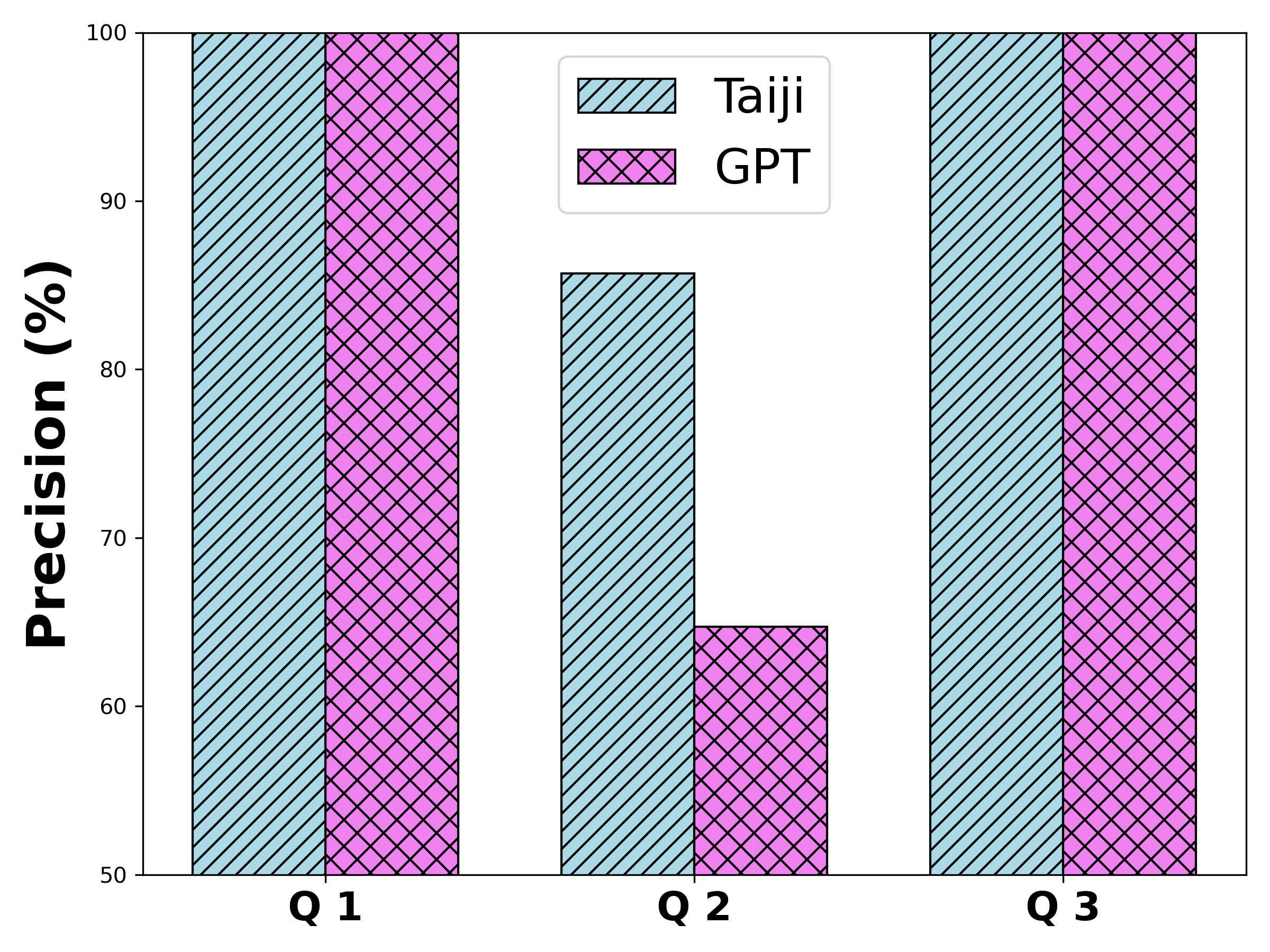}
    \captionof{figure}{Precision Evaluation.}
    \label{fig:precision}
\end{minipage}
\hfill
\begin{minipage}[t]{0.3\textwidth}
	\centering
	\includegraphics[width=\textwidth]{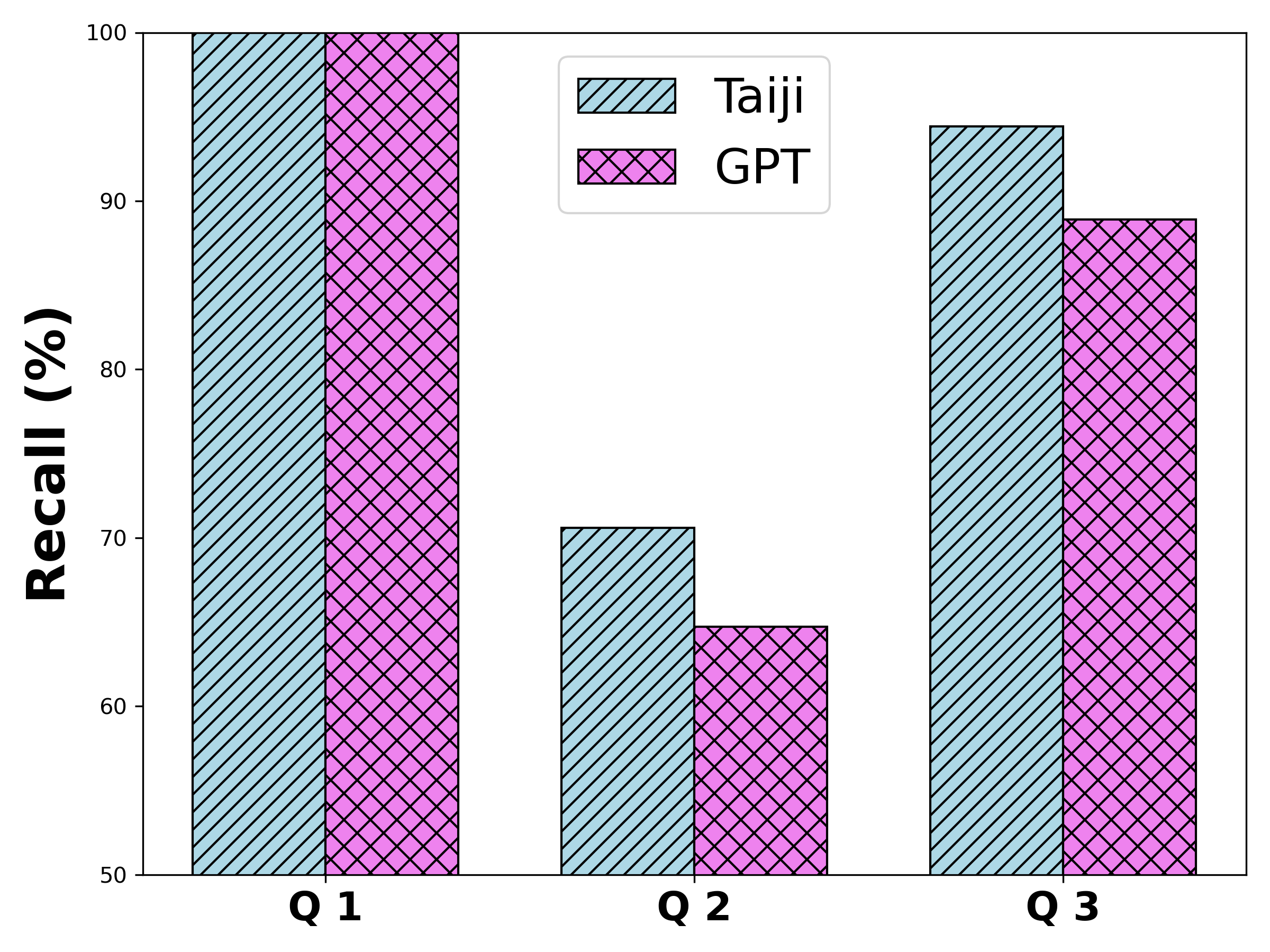}
	\captionof{figure}{Recall Evaluation.}
	\label{fig:recall}
\end{minipage}
\hfill
\begin{minipage}[t]{0.3\textwidth}
    \centering
    \includegraphics[width=\textwidth]{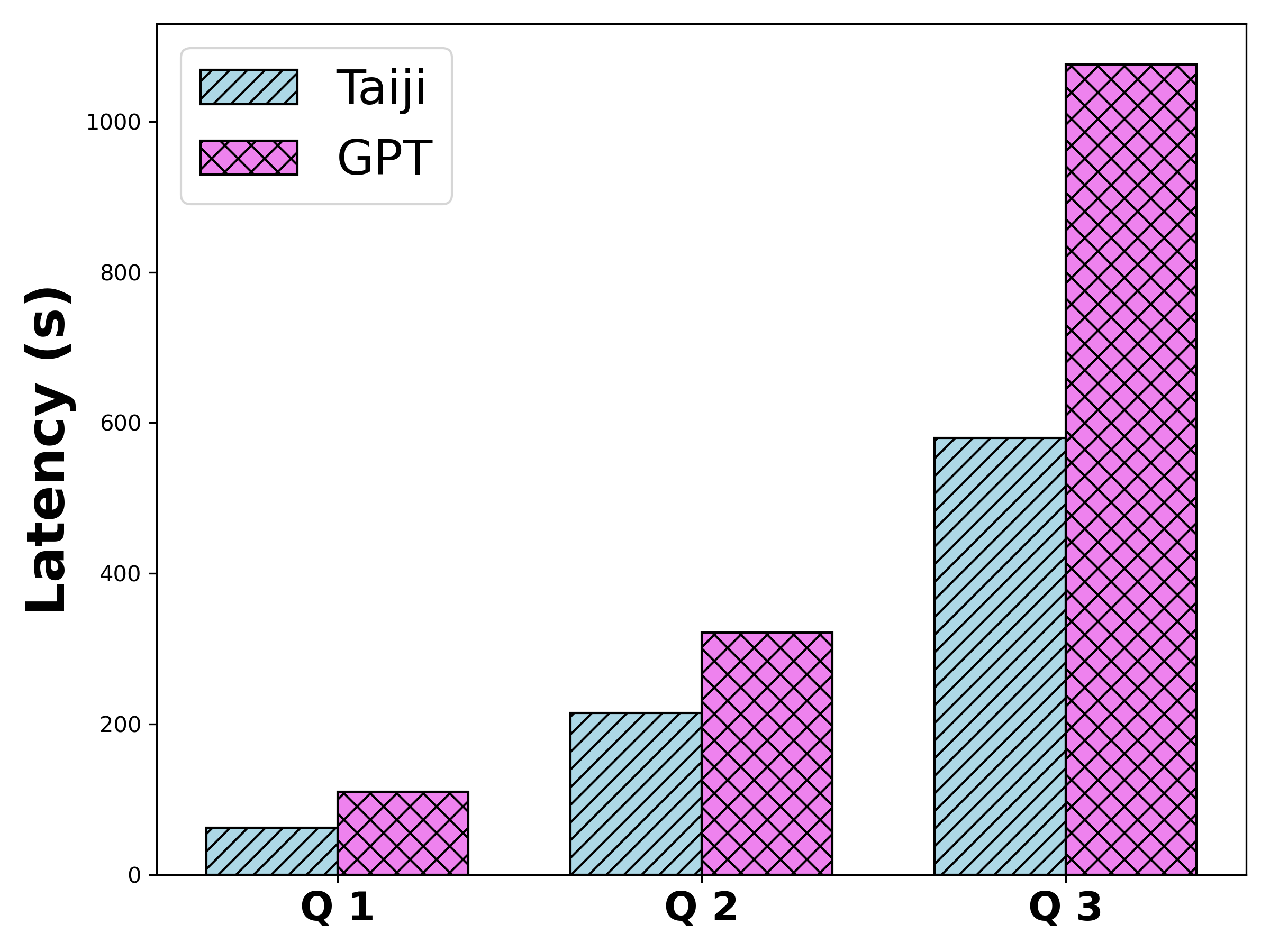}
    \captionof{figure}{ Latency Evaluation.}
    \label{fig:latency}
\end{minipage}%
\vspace{1em}

% Taiji_latency = [62.32, 214.66, 580.29]
% GPT_latency = [110.02, 321.40, 1076.01]
% Taiji_recall = [100.00, 70.59, 94.44]
% GPT_recall = [100.00, 64.71, 88.89]
% Taiji_acc = [100.00, 85.71, 100.00]
% GPT_acc = [100.00, 64.71, 100.00]

As shown in Figure \ref{fig:precision} and Figure \ref{fig:recall}, \ours also outperforms GPT in accuracy (for Q2) and recall (on Q2 and Q3). In specific, \ours achieves an accuracy of 85\%  on Q2 while GPT has an accuracy of 65\%.; \ours has a recall of 71\% and 94\% on Q2 and Q3, respectively while GPT has a s a recall of 65\% and 89\%. Take Q2 as an example where \ours is better than GPT both in precision and recall, we found that it is hard to find the black chair from the candidate images, the reason is that GPT can hardly figure out black chair and black sofa. 

Figure \ref{fig:latency} depicts the latency between \ours and GPT-4.1. It is clearly visible that our approach is largely faster than GPT. As the intermediate results increase (i.e., Q1-Q3), the performance gap increases. On average, \ours improves the latency by 43\%. Take Q3 as an example where the intermediate size has reached up to 347, GPT is too heavy to offer high efficiency with its trillion parameters.

\textbf{Implications of MCP-based Architecture over pure LLM.} Through the preliminary experiments, we have the following insights. First, compared with pure LLM, MCP-based architecture can judiciously select and harness most proper LLMs to process multi-modal data. For instance, GPT may be excel at text processing and query translation, but it turns out that Qwen is better than GPT on image processing. Second, leveraging a tailored LLM (i.e., Qwen) with a much smaller number of parameters can not only improve the accuracy, but also it can result in a much lower inference overhead. This implies that, MCP-based  architecture is very promising to combine many small LLMs for a wide range of data modality. Third, since MCP-based architecture can offload a sub-task to a local MCP server, it further reduces the burden of the centric LLM. Such an observation motivates us to explore more optimizations in the future, such as asynchronous execution, task scheduling, etc.

% \fanj{more in-depth analysis.}

\section{Challenges and Opportunity}

\noindent  \textbf{MCP-based Data Security.} Despite the salient features discussed in the previous sections, a major challenge of MCP-based data analytic is how to preserve the data security while performing the LLM-empowered data analytics. This is because MCP servers are developed by different vendors and contributors all over the world, and analyzing the plain text/documents in the data lake poses high risks of data leakage. For instance, the MCP SDK or build-in LLM agents may upload the data to the web server. Therefore, it calls for new mechanisms to enable privacy-preserved data transmission and analytics. On the one hand, the data transmission tunnel should be secured among MCP host/clients, MCP servers, and resources. On the other hand, it shoud avoid accessing the plain text, other alternatives should be explored,  such as query processing over ciphertext directly. However, it is challenging to balance the trade-off between query efficiency and data privacy.

\noindent  \textbf{Hybrid Deployment Optimization.} In addition to the local deployment of MCP servers, it is also possible to deploy remote MCP servers, resulting in a hybrid deployment. Hence, it calls for new methods that can harness the capability of hyrid deployment. However, it is challenging to effectively determine the strategy of data placement and to migrate the data among the multiple MCP servers. Moreover, it is also hard to efficiently discover data of high quality due to the hugh search space and high latency.

\noindent  \textbf{Cost-Aware Model Mangement.}  As more and more MCP servers are involved, it has become a challenge to manage the multiple LLMs and their evolvement. The reason is two-fold. First, existing approaches~\cite{urban2024eleet,jo2024thalamusdb,wang2025aop} only care about the query performance, while neglecting the dollar cost of LLMs' invocation as they charge for tokens. Second, model knowledge can also be obsolete as discussed before, and performing the fine-tuning also incur significant cost, thus it is challenging to balance the knowledge freshness and training cost. As a result, it is required to judiciously select and fine-tune the LLM models with cost-aware optimization.

% Existing cloud-native databases are either OLTP-oriented systems or OLAP-oriented systems, and there are no cloud-native HTAP systems[9,12]. The main challenge is how to judiciously schedule the resources for OLTP and OLAP workloads with SLA-aware optimization.

\noindent \textbf{Multi-Modal Database Benchmark.} Since there lacks a unified and standard multi-modal database benchmark, existing approaches basically evaluate the performance on different datasets and workloads. Hence, there is a pressing need to have a new multi-modal database benchmark that covers various data modality and  workloads. However, it is challenging to collect or generate multi-modal data and workloads with both fidelity, utility, and generality, even for one modality, there could be numerous variants. 

%The evaluation of LLM-driven analytics over multi-modal data lakes remains an open challenge, particularly in comparing the effectiveness of different methodologies. Metrics for traditional data analytics, such as precision, recall, or F1 score, may not adequately capture the nuances of integrating structured and unstructured data sources. Furthermore, due to the large-scale and dynamic nature of the data lake, it is difficult to develop comprehensive benchmarks that can consistently evaluate the model’s performance across different types of queries and use cases. Developing new benchmarking methodologies tailored for multi-modal data lake analytics would be an essential step forward. These benchmarks could focus on both the accuracy of the results and the system’s ability to integrate and analyze heterogeneous data sources. Additionally, creating standardized test datasets for data lake environments could encourage further research and provide a basis for consistently comparing different models and systems.

\section{Conclusion}
\label{sec:conclusion}

%In this paper, we present \ours, a novel LLM-powered analytics system designed to address
%the challenges of data analytics in multi-modal data lakes. By integrating large language models with advanced semantic operators, embedding-based data linking, and dynamic pipeline orchestration, \ours provides a unified framework for querying and analyzing diverse data types, including structured, semi-structured, and unstructured data. iDataLake introduces several key innovations including semantic operators tailored to the unique requirements of multi-modal analytics, unified embedding-based data linking for aligning heterogeneous data types in a common semantic space, dynamic pipeline adjustment to adapt to evolving query execution requirements and interactive and incremental plan execution to ensure robust and efficient query handling.

In this paper, we present \ours,  a novel multi-modal data analytical system. Particularly, we design a new architecture based on Model Context Protocol (MCP) that offloads the multi-modal operators to specific MCP server. We propose a series of components including NL2Operator query translation, multi-modal query planning, iterative RAG, embedding indexing, data augmentation, model refreshing. To the best of our knowledge, this is the first MCP-based architecture for multi-modal analytical system. We believe MCP architecture opens the door for the researches over multi-modal data management. In the future, we will finish implementing the whole system and conduct more experiments.

\small
%\section*{Acknowledgments}

\bibliographystyle{abbrv}%{ACM-Reference-Format}
\bibliography{unify}

\end{document}